# Variation of the physical properties of four transition metal oxides SrTMO$_3$ (TM = Rh, Ti, Mo, Zr) under pressure: An *ab initio* study

Md. Lokman Ali[1], Md. Zahidur Rahaman[2]

[1, 2]*Department of Physics, Pabna University of Science and Technology, Pabna-6600, Bangladesh*

**Abstract**

Using first principles calculations, we have studied the structural and elastic properties of SrTMO$_3$ (TM = Rh, Ti, Mo, Zr) under pressure by the plane wave pseudopotential method based on the density functional theory (DFT) within the generalized gradient approximation (GGA). The optical properties have been investigated under zero pressure. It is observed that the calculated lattice parameters are in good agreement with the available experimental result and previous theoretical results. From the static finite strain method, the independent elastic constants and their pressure dependence are calculated. The effect of pressures, up to 60 GPa, on the lattice parameters, bulk modulus *B*, Shear modulus *G*, *B/G*, Poisson's ratio *v*, anisotropy factor *A* are also investigated. This is the first theoretical prediction of the elastic and optical properties of SrTMO$_3$ compounds under pressure. All these calculations have been carried out using the CASTEP computer code.

## 1. Introduction

Transition metal oxides (TMO) have been attractive to research community due to their remarkable magnetic, electronic and transport properties [1-3]. Transition metal oxides are widely used in the technological applications such as catalysis, microelectronics, substrates for growth of high T$_c$ superconductors, gas sensors, and thin films of cubic SrTMO$_3$ perovskite [4]. The TMO family is the most widely studied material due to their cubic crystal structure at room temperature and high dielectric constant. A number of theoretical and experimental works have been carried out on the structural, electronic and elastic properties of SrTMO$_3$ compounds at zero pressure. I.R. Shein *et al.* [5] have performed a density functional study on structural, elastic and electronic properties of cubic SrMO$_3$ (M = Ti, V, Zr and Nb) crystal. A.J. Smith *et al.* [6] has been performed a theoretical investigation on the phase stability, structural, elastic and electronic properties of some mixed metal oxides.

It is thus evident from the above discussion that the majority of the cited works discussed briefly the electronic properties and only a few of them discussed the elastic properties at zero pressure of perovskite crystals. To the best of our knowledge pressure effects on structural and elastic properties with the optical properties at zero pressure of SrTMO$_3$ (where TM = Rh, Ti, Mo, Zr) compounds have not yet been discussed in literature. Thus, in the present work, we focus on the pressure effects of the structural and elastic properties of SrTMO$_3$-type perovskites where TM stands for Rh, Ti, Mo, and Zr respectively by using the density functional theory (DFT) method as implemented in the CASTEP code. The format of this paper is organized as follows: the computational methods are given in section 2, results and discussion are presented in section 3. Finally, the summery of our main results is given in section 4.

………………………………………………………………………………………….
[1]Corresponding Author: lokman.cu12@gmail.com

## 2. Computational methods and details

The computations have been performed by using the first principles method based on the density functional theory [7] as implemented in the CASTEP code [8]. Generalized gradient approximation (GGA) of Perdue-Burke-Ernzerhof (PBE) [9] is used for the exchange and correlation energy function. The ultrasoft pseudo-potential [10] is used to describe the interaction between ion core and valence electron. In the present calculations, the cut-off energy of plane wave is set to 340 eV, we get a good convergence using $6 \times 6 \times 6$ set of Monkhorst-Pack mesh [11] grid for the total energy calculation. The optimizations of structural parameter were conducted by using the Broyden-Fletcher-Goldfarb-Shanno (BFGS) minimization [12]. In the geometry optimization criteria of convergence were set as follows: i) the difference in total energy is set to $1.0 \times 10^{-5}$ eV/atom, ii) the maximum ionic force is set to 0.03 eV/ Å, iii) the maximum ionic displacement is set to $1.0 \times 10^{-3}$ Å, iv) the maximum stress is set to 0.05 GPa. Thus, the present parameters are sufficient to lead to a good converged total energy.

The elastic constants of $SrTMO_3$ compounds are obtained by the stress-strain method [13] at the optimized structure under the condition of each pressure. In this method we set the tolerances within i) $1.0 \times 10^{-7}$ eV/atom for energy ii) within $6.0 \times 10^{-4}$ eV/Å for the maximum force, iii) within $2.0 \times 10^{-4}$ Å for the maximum ionic displacement and iv) within 0.003 GPa for the maximum strain amplitude in the present calculation.

## 3. Results and discussion

### 3.1. Structural properties

Transition metal oxides $SrTMO_3$ (TM = Rh, Ti, Mo, Zr) belongs to cubic crystal structure of space group Pm-3m (221). The value of the equilibrium lattice parameters of $SrTMO_3$ (TM = Rh, Ti, Mo, Zr) are shown in Table 1. We have optimized the lattice parameters and atomic positions of these compounds as a function of normal stress by minimizing the total energy. The optimized crystal structure of $SrTMO_3$ (TM = Rh, Ti, Mo, Zr) is illustrated in Fig. 1. The evaluated values of the structural properties of $SrTMO_3$ (TM = Rh, Ti, Mo, Zr) at zero pressure are tabulated in Table 1 with others experimental and theoretical values. It is evident from Table 1 that our calculated values match well with the experimental values as well as other theoretical values indicating the reliability of our present DFT based investigation. The evaluated lattice constants of $SrRhO_3$, $SrTiO_3$, $SrMoO_3$ and $SrZrO_3$ are respectively 4.074, 3.957, 4.003 and 4.177 Å which shows small deviation from the experimental values. The different condition and calculation method may be the reason for the existing discrepancy.

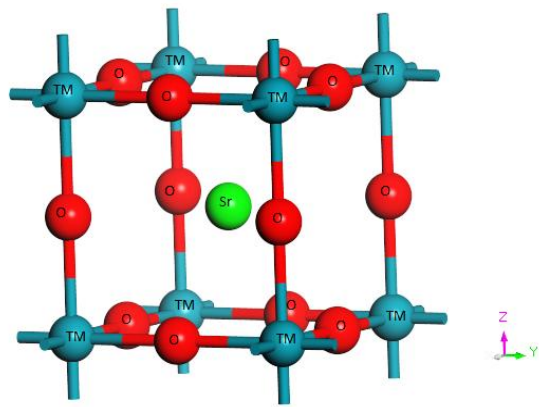

**Fig. 1.** The crystal structure of $SrTMO_3$ (TM = Rh, Ti, Mo, Zr) compounds.

**Table 1.** Calculated structural parameters of SrTMO$_3$ (TM = Rh, Ti, Mo, Zr) at P = 0 GPa.

| Materials | Referance | a (Å) | V (Å$^3$) |
|---|---|---|---|
| SrRhO$_3$ | This work | 4.074 | 67.63 |
| | Experimental [14] | 3.970 | -- |
| | Theoretical [14] | 3.920 | -- |
| SrTiO$_3$ | This work | 3.957 | 61.98 |
| | Experimental [15] | 3.905 | -- |
| | Theoretical [14] | 3.878 | -- |
| SrMoO$_3$ | This work | 4.003 | 63.04 |
| | Experimental [14] | 3.980 | -- |
| | Theoretical [16] | 3.994 | -- |
| SrZrO$_3$ | This work | 4.177 | 72.92 |
| | Experimental [6] | 4.109 | -- |
| | Theoretical [14] | 4.095 | -- |

To study the influence of external stress on the structure of SrTMO$_3$ (TM = Rh, Ti, Mo, Zr), the variations of the lattice parameters and unit cell volume of SrTMO$_3$ (TM = Rh, Ti, Mo, Zr) with different pressure up to 60 GPa have been investigated and shown in Fig. 2. It is evident from Fig. 2 that both the ratio $a/a_0$ and normalized volume $V/V_0$ decreases with the increase of pressure, where $a_0$ and $V_0$ are defined as equilibrium lattice parameter and volume at zero pressure respectively. However it is noticed that the interatomic distance is reduced with the increase of pressure. The pressure-volume curves of SrTMO$_3$ (TM = Rh, Ti, Mo, Zr) transition metal oxides are illustrated in Fig. 2(b). It is clear from Fig. 2(b) that volume of SrTMO$_3$ (TM = Rh, Ti, Mo, Zr) metal oxides are decreased with the increase of pressure. The achieved pressure-volume data of SrTMO$_3$ (TM = Rh, Ti, Mo, Zr) are fitted well to a third-order Birch-Murnaghan equation of state (EOS) [17].

**Table 2.** Calculated structural parameters of (a) SrRhO$_3$ (b) SrTiO$_3$ (c) SrMoO$_3$ (d) SrZrO$_3$ at different hydrostatic pressures.

| Materials | P (GPa) | a (Å) | V (Å$^3$) |
|---|---|---|---|
| SrRhO$_3$ | 20 | 3.935 | 60.95 |
| | 40 | 3.840 | 56.65 |
| | 60 | 2.770 | 53.61 |
| SrTiO$_3$ | 20 | 3.826 | 56.03 |
| | 40 | 3.736 | 52.18 |
| | 60 | 3.668 | 49.36 |
| SrMoO$_3$ | 20 | 3.885 | 58.66 |
| | 40 | 3.796 | 54.71 |
| | 60 | 3.733 | 52.05 |
| SrZrO$_3$ | 20 | 4.032 | 65.55 |
| | 40 | 3.931 | 60.78 |
| | 60 | 3.855 | 57.33 |

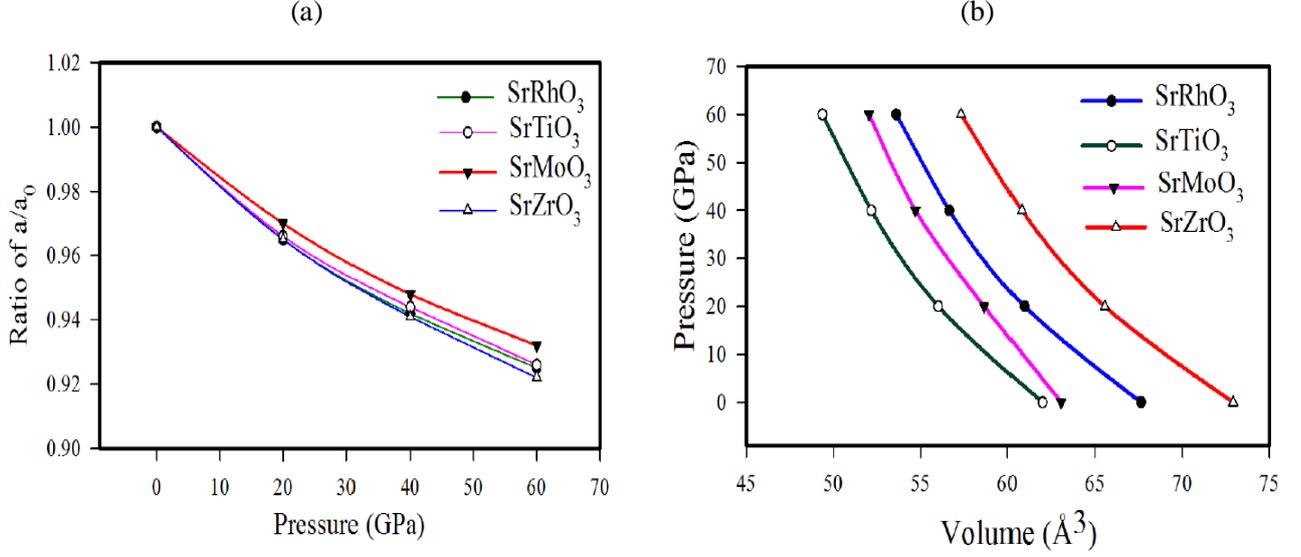

**Fig. 2.** Variation of lattice parameters as function of pressure (a). Birch-Murnaghan equation of state for SrTMO$_3$ (TM = Rh, Ti, Mo, Zr) (b).

*3.2 Elastic Properties*

The elastic constants are significant parameters of solids. They are related to fundamental solid state phenomena such as brittleness, ductility, stiffness of materials and the mechanical stability of materials [18]. Elastic constants provide useful information concerning the structural stabilities, bonding behaviour, anisotropic factor and the cohesion of material [19]. The elastic constants of materials under pressure are essential to understand the resistance of a crystal to an externally applied different pressure. Hence, it is significant to investigate the mechanical properties of SrTMO$_3$ (TM = Rh, Ti, Mo, Zr) compounds to an externally applied stress. The SrTMO$_3$ compounds belong to the cubic structure which has three independent elastic constants such as $C_{11}$, $C_{12}$ and $C_{44}$. In this work, the calculated elastic constants of SrTMO$_3$ compounds at pressure up to 0 GPa to 60 GPa with a step of 20 GPa for the first time.

In Table 3, we listed the elastic constants of SrTMO$_3$ compounds at zero pressure. For a cubic crystal, the criterion for mechanical stability is [20],

$$C_{11} > 0,\ C_{44} > 0,\ C_{11} - C_{12} > 0 \text{ and } C_{11} + 2C_{12} > 0 \quad (1)$$

Our calculated data satisfy all these above conditions up to 60 GPa indicating that SrTMO$_3$ compounds is mechanical stable. To the best of our knowledge, there is no experimental result for our comparison, so we compare the theoretical data of SrTMO$_3$ compounds at 0 GPa pressure [6] as reference. From Table 3, it can be seen that our investigated data are good consistent with theoretical data. In Table 3 and Table 4, we present our calculated elastic constants ($C_{11}$, $C_{12}$ and $C_{44}$), bulk modulus $B$, shear modulus $G$, $B/G$, Young's modulus $E$, Poisson's ratio $v$, and anisotropic factor $A$ of the cubic SrTMO$_3$ at 0 GPa and up to 60 GPa with a step of 20 GPa. The pressure dependence of elastic parameters is described in Fig. 3 and Fig. 4. It is noticed that the elastic constants $C_{11}$, $C_{12}$ and $C_{44}$ almost increase monotonically with the applied pressure and are slightly sensitive to the variation of pressure of all the compounds except SrMoO$_3$ compound. As is shown $C_{44}$ of SrMoO$_3$ compound almost remain unchanged and applied pressure 20 GPa to 40 GPa at SrMoO$_3$ compound the elastic constant $C_{11}$ remain unchanged. Elastic constants are used to determine the mechanical properties of materials such as Young's modulus, shear modulus, Poisson's ratio and anisotropy factor for useful application. For the cubic crystal, the Voigt bounds [21] and Reuss bound [22] of the bulk modulus and shear modulus are given as:

$$B_v = B_R = \frac{(C_{11} + 2C_{12})}{3} \quad (2)$$

$$G_v = \frac{(C_{11} - C_{12} + 3C_{44})}{5} \quad (3)$$

The Reuss bounds of the bulk modulus and shear modulus are:

$$B_v = G_v \quad (4)$$

$$\text{and,} \quad G_R = \frac{5C_{44}(C_{11} - C_{12})}{[4C_{44} + 3(C_{11} - C_{12})]} \quad (5)$$

The expression of bulk modulus $B$ and shear modulus $G$ are given as follows:

$$B = \frac{1}{2}(B_R + B_v) \quad (6)$$

$$G = \frac{1}{2}(G_v + G_R) \quad (7)$$

Using the bulk modulus $B$ and shear modulus $G$, the Young's modulus $E$ and Poisson's ratio ($v$) and anisotropic factor ($A$) are obtained according to the following formula [23]

$$E = \frac{9GB}{3B + G} \quad (8)$$

$$v = \frac{3B - 2G}{2(3B + G)} \quad (9)$$

$$A = \frac{2C_{44}}{(C_{11} - C_{12})} \quad (10)$$

The computed values of the bulk modulus $B$, shear modulus $G$, Young's modulus $E$, Poisson's ratio $v$ and anisotropic factor $A$ are presented in Table 3.

**Table 3.** Calculated elastic constants $C_{ij}$ (GPa) of SrTMO$_3$ (TM = Rh, Ti, Mo, Zr) at P = 0 GPa.

| Materials | Referance | $C_{11}$ | $C_{12}$ | $C_{44}$ | B | G | E | B/G | $v$ | A |
|---|---|---|---|---|---|---|---|---|---|---|
| SrRhO$_3$ | Present | 196.25 | 99.90 | 46.28 | 132.01 | 47.02 | 126.08 | 2.80 | 0.34 | 0.97 |
| | Expt. | -- | -- | -- | -- | -- | -- | -- | -- | -- |
| | Theory | -- | -- | -- | -- | -- | -- | -- | -- | -- |
| SrTiO$_3$ | Present | 273.46 | 85.18 | 96.90 | 147.94 | 95.78 | 236.33 | 1.54 | 0.23 | 1.02 |
| | Expt.[25] | -- | -- | 155.00 | 184.00 | -- | -- | -- | -- | -- |
| | Theory[24] | -- | -- | -- | 200.00 | 109.16 | -- | -- | -- | -- |
| SrMoO$_3$ | Present | 336.61 | 76.76 | 56.95 | 163.37 | 79.79 | 205.85 | 2.04 | 0.28 | 0.43 |
| | Expt. | -- | -- | -- | -- | -- | -- | -- | -- | -- |
| | Theory | -- | -- | -- | -- | -- | -- | -- | -- | -- |
| SrZrO$_3$ | Present | 299.16 | 72.57 | 72.58 | 124.76 | 74.81 | 187.04 | 1.66 | 0.25 | 0.92 |
| | Expt. | -- | -- | -- | -- | -- | -- | -- | -- | -- |
| | Theory [26] | 338.60 | 71.00 | 77.00 | 160.00 | 118.88 | -- | 1.34 | 0.19 | -- |

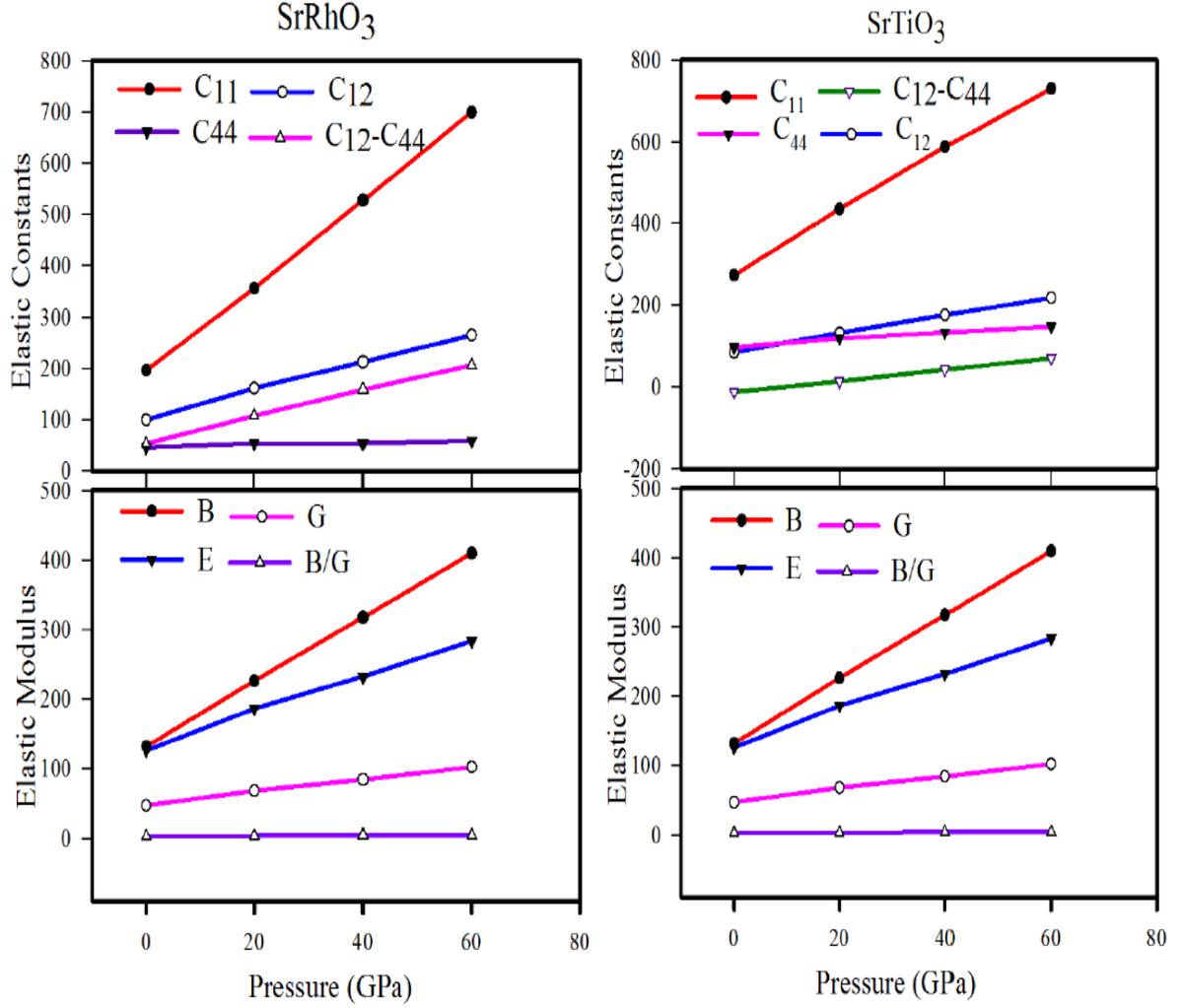

**Fig. 3.** The calculated elastic constants and elastic parameters ($B$, $G$, $E$, $B/G$) of SrRhO$_3$ and SrTiO$_3$ compounds under different pressure.

**Table 4.** The calculated elastic constants, bulk modulus $B$ (GPa), shear modulus $G$ (GPa), Young's modulus $E$ (GPa), $B/G$ values, Poisson's ratio $v$ and anisotropy factor $A$ of SrTMO$_3$ (M = Rh, Ti, Mo, Zr) compounds under hydrostatic pressure.

| Materials | $P$ (GPa) | $C_{11}$ | $C_{12}$ | $C_{44}$ | $B$ | $G$ | $E$ | $B/G$ | $v$ | $A$ |
|---|---|---|---|---|---|---|---|---|---|---|
| SrRhO$_3$ | 20 | 356.51 | 161.34 | 53.70 | 226.39 | 68.36 | 186.32 | 3.31 | 0.36 | 0.55 |
|  | 40 | 527.60 | 212.57 | 54.12 | 317.58 | 84.42 | 232.64 | 3.76 | 0.37 | 0.34 |
|  | 60 | 699.90 | 264.97 | 58.53 | 409.94 | 102.40 | 283.58 | 4.00 | 0.38 | 0.26 |
| SrTiO$_3$ | 20 | 435.43 | 132.65 | 118.81 | 233.57 | 130.92 | 330.92 | 1.78 | 0.26 | 0.78 |
|  | 40 | 587.99 | 176.01 | 133.55 | 313.33 | 158.96 | 407.90 | 1.97 | 0.28 | 0.64 |
|  | 60 | 729.84 | 218.41 | 147.76 | 388.88 | 184.36 | 477.60 | 2.10 | 0.29 | 0.57 |
| SrMoO$_3$ | 20 | 537.89 | 123.58 | 74.74 | 261.68 | 114.05 | 298.74 | 2.29 | 0.30 | 0.36 |
|  | 40 | 518.39 | 247.72 | 72.05 | 337.94 | 92.99 | 255.53 | 3.63 | 0.37 | 0.53 |
|  | 60 | 638.56 | 292.34 | 64.99 | 407.74 | 97.43 | 270.72 | 4.18 | 0.39 | 0.37 |
| SrZrO$_3$ | 20 | 461.75 | 104.35 | 72.16 | 223.48 | 104.76 | 271.80 | 2.13 | 0.29 | 0.40 |
|  | 40 | 625.15 | 147.40 | 87.14 | 306.65 | 116.82 | 310.97 | 2.62 | 0.33 | 0.36 |
|  | 60 | 745.87 | 176.00 | 86.55 | 365.95 | 142.92 | 379.37 | 2.56 | 0.32 | 0.30 |

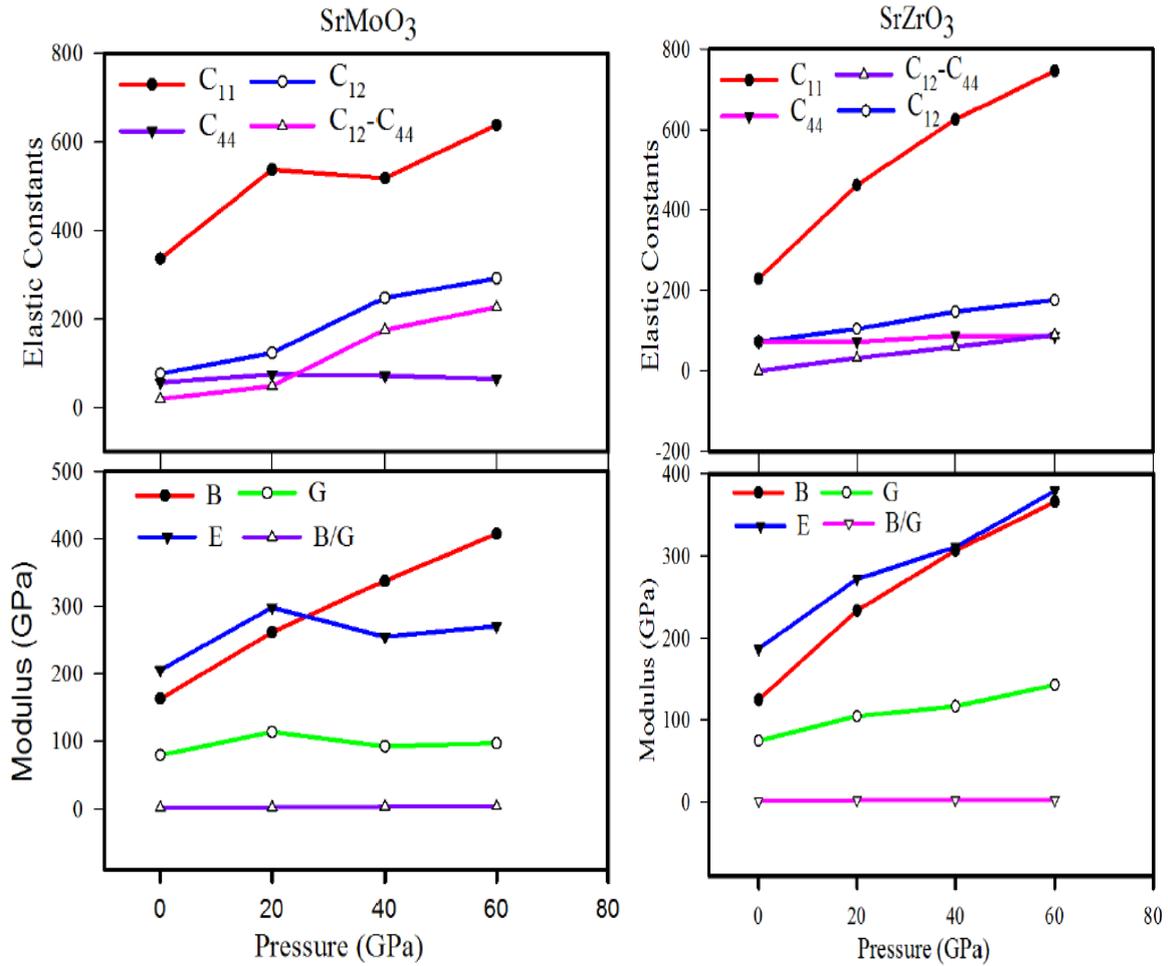

**Fig. 4.** The calculated elastic constants and elastic parameters (*B, G, E, B/G*) of SrMoO$_3$ and SrZrO$_3$ compounds under different pressure.

The bulk modulus or shear modulus can measure the hardness of the material [27]. From Table 4, it can be seen that the bulk modulus increases with the increase in the applied pressure, noticed that the cubic structure of SrTMO$_3$ becomes more strength to compress with the increasing pressure.

Young's modulus (*E*) is an important parameter for technological and engineering application, which is defined as the ratio of stress and strain, and is to provide a measure of the stiffness of the material. The highest value of E, the stiffer is the material and the stiffer solids have covalent bonds. From Table 4, it can be seen that the highest value of Young's modulus of SrTMO$_3$ compound is indicating the presence of covalent bonds. In Fig. 3 and Fig. 4, we also displayed the pressure dependence of Young's modulus of SrTMO$_3$ compounds. It is shown that the Young's modulus has an increasing trend with the increasing pressure.

Pugh [28] reported that the ratio of *B/G* to distinguish the ductility and brittleness of the materials. If the value of *B/G* is greater than 1.75, a material behaves in a ductile nature; otherwise it behaves in a brittle nature. In our present work, the *B/G* ratio of SrTiO$_3$ and SrZrO$_3$ is 1.54 and 1.66 at ground state, indicating that it is brittle manner. The *B/G* ratio of SrRhO$_3$ and SrMoO$_3$ compounds is 2.80 and 2.04 respectively at zero pressure indicating that it is ductile. From Fig. 3 and Fig. 4, it can be found that the *B/G* ratio increases with the increasing pressure, indicating that it becomes more ductile as the pressure increases.

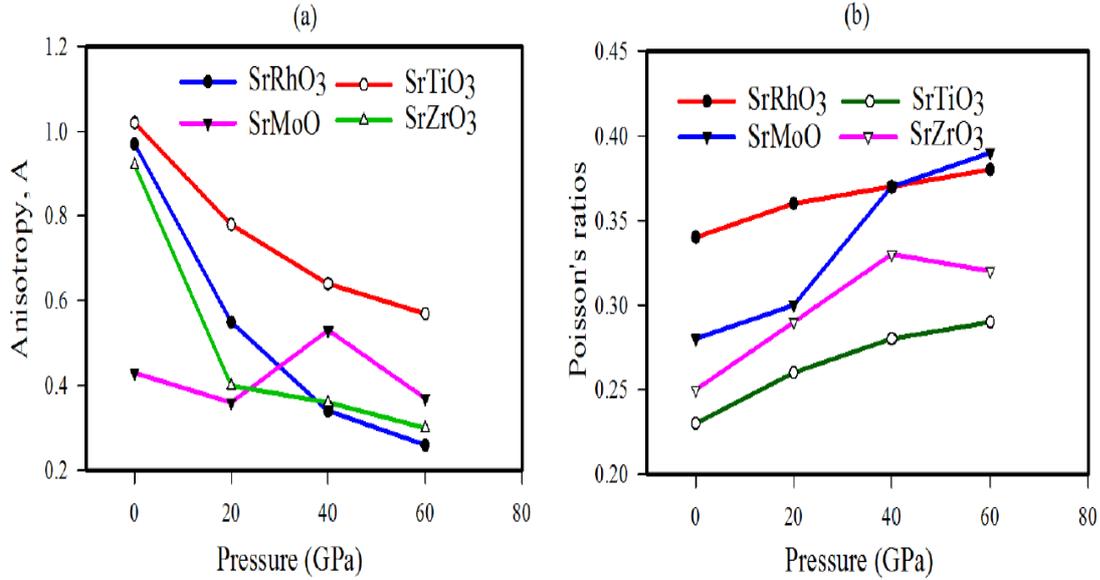

**Fig. 5**: The calculated Anisotropy factor *A* and Poisson's ratio *v* of SrTMO$_3$ (TM = Rh, Ti, Mo, Zr) compounds under various pressure.

The value of Poisson's ratio is used to measure the stability of the material and provides useful information about the nature of the bonding forces [29]. According to the Frantsevich [30] rule, the critical value of Poisson's ratio is greater than 1/3 the material behaves in ductile nature, otherwise the materials behaves in a brittle manner. At ground state, the highest value of Poisson's ratio for SrRhO$_3$ reveals that this is most ductile amongst of all SrTMO$_3$ compounds. The effect of pressure of Poisson's ratio is also displayed in Fig. 5. It can be seen that the value of increases with the increase of pressure.

The elastic anisotropy is an important parameter to measure of the degree of anisotropy of materials [31]. Anisotropy factor has important implication in material science. In Table 4, we also present the calculated anisotropy factor from elastic constants. For an isotropic material, the value of *A* is unity otherwise; the material has an elastic anisotropy [32]. For compounds (SrRhO$_3$, SrMoO$_3$ and SrZrO$_3$) it is found less than unity, while for compound SrTiO$_3$ it is found greater than unity. From Table 4, it can be seen that the SrTMO$_3$ compounds is elastically anisotropic crystal. In Fig.5 displayed the effect of pressure on the anisotropy factor. It is shown that the anisotropy factor have a decreasing trend with the increasing pressure, which suggest that the degree of elastic anisotropy is getting smaller.

*3.3 Optical properties*

The study of the optical properties of material is important for comprehending the electronic structure and other physical properties preciously. In this section we have discussed about the details optical properties of four transition metal oxides SrTMO$_3$ (TM = Rh, Ti, Mo, Zr). The investigated optical parameters such as the refractive index, dielectric function, the energy loss function, the absorption spectrum, the optical conductivity and the reflectivity of transition metal oxides SrTMO$_3$ (TM = Rh, Ti, Mo, Zr) for incident light energy up to 80 eV are depicted in Fig. 6 to Fig.8. These optical parameters had been determined by using the complex dielectric function *ε (ω)* which can be written as *ε (ω) = ε$_1$ (ω) + iε$_2$ (ω)*, where *ε$_1$ (ω)* and *ε$_2$ (ω)* are the real and imaginary part respectively.

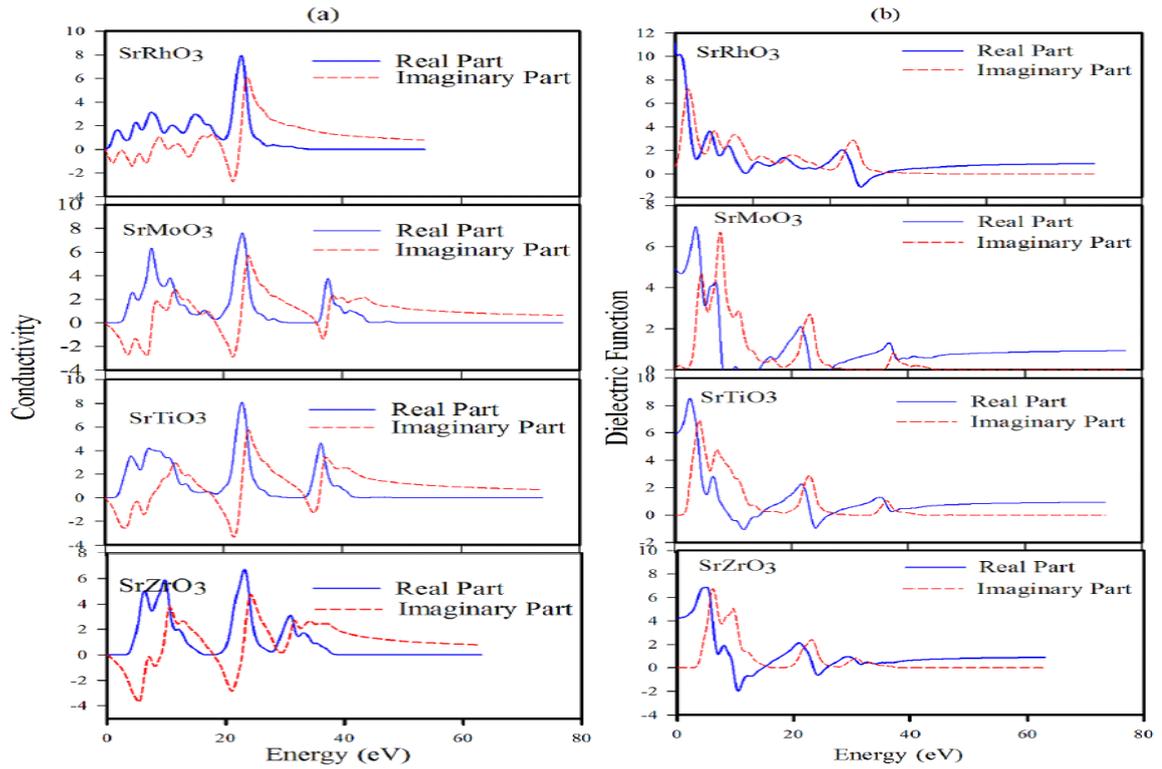

**Fig.6.** The conductivity (a), and dielectric function (b) of SrTMO$_3$ (TM = Rh, Ti, Mo, Zr).

Fig. 6(a) illustrates the conductivity spectra of SrTMO$_3$ (TM = Rh, Ti, Mo, Zr) metal oxides as a function of photon energy up to 80 eV for the polarization vector [100]. According to the plot the photoconductivity for all of the four metal oxides begin with zero photon energy indicating the compounds under investigation have zero band gaps. We have also observed several maxima and minima with similar peak for all the compounds in the conductivity plot. However, electrical conductivity of SrTMO$_3$ (TM = Rh, Ti, Mo, Zr) raises as a result of absorbing photon energy [33].

The dielectric function is an important optical parameter which is used to characterize the response of a material to the incident beam of radiation (electromagnetic wave). In Fig. 6(b) the real and imaginary parts of dielectric function are illustrated of SrTMO$_3$ (TM = Rh, Ti, Mo, Zr) to the polarization vector [100]. It can be seen from the figure that $\varepsilon_2$ (imaginary part) becomes zero at about 40 eV, 42 eV, 41 eV, and 38 eV for SrRhO$_3$, SrMoO$_3$, SrTiO$_3$ and SrZrO$_3$ respectively which indicating that these materials become transparent above these certain values. It is also clear that all of the four metal oxides exhibit minor absorption within the energy range up to 80 eV since for nonzero value of $\varepsilon_2(\omega)$ absorption occurs [34]. The static dielectric constants of SrRhO$_3$, SrMoO$_3$, SrTiO$_3$ and SrZrO$_3$ are 10, 5, 6 and 4 respectively. Hence from higher to lower according to the value of the dielectric constant we can arrange these four metal oxides as SrRhO$_3$> SrTiO$_3$ > SrMoO$_3$> SrZrO$_3$ since metals with high dielectric constant is used for manufacturing high value capacitors [35]

The absorption coefficient spectra of SrTMO$_3$ (TM = Rh, Ti, Mo, Zr) is depicted in Fig. 7(c) according to which the spectra for all the four oxides begin from 0 eV indicating the metallic nature of these four compounds. We observed several picks in the plot with the highest peak located at about 25 eV for all the four metals. All these four oxides possess good absorption coefficient from 0 eV to 43 eV.

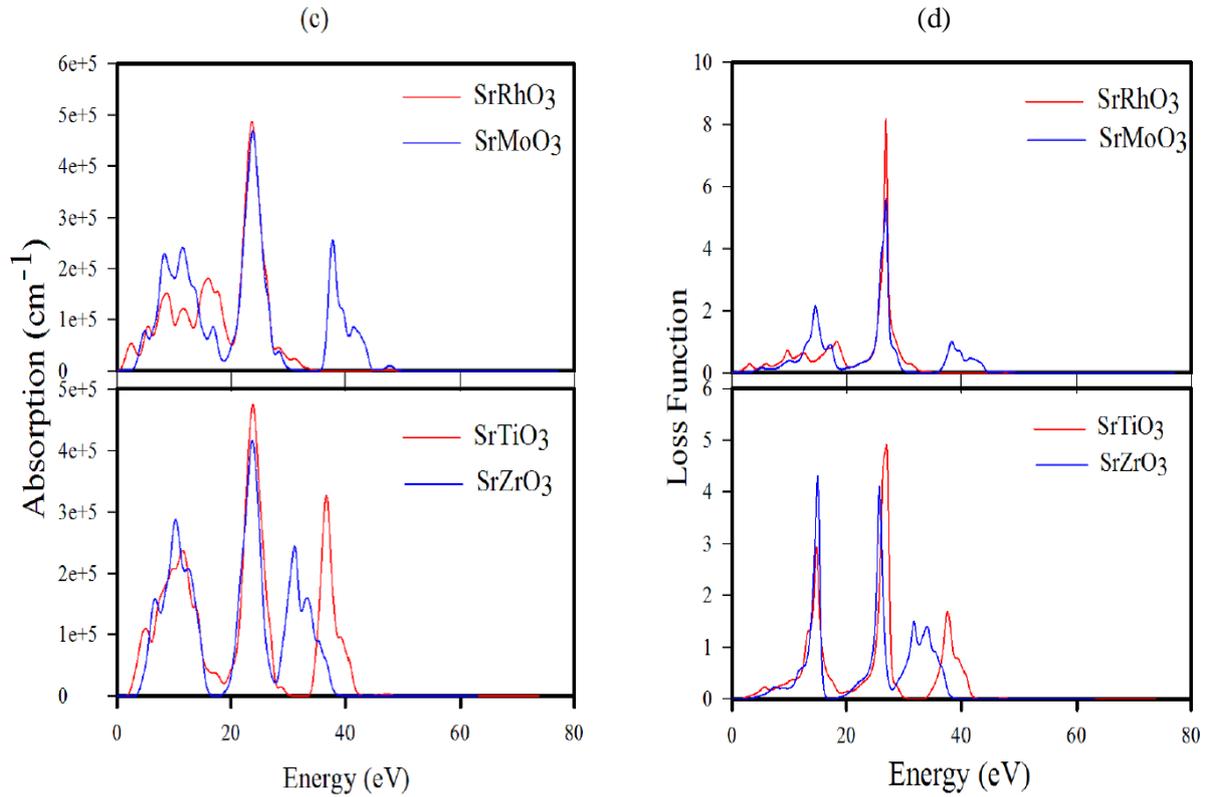

**Fig. 7.** The Absorption coefficient (c), and Loss function (d) of $SrTMO_3$ (TM = Rh, Ti, Mo, Zr) metal oxides.

The energy loss spectra of $SrTMO_3$ (TM = Rh, Ti, Mo, Zr) are plotted in Fig. 7(d). It is a crucial optical parameter which is used for comprehending the screened excitation spectra produced by the swift charges inside a material. From the figure the bulk plasma frequency (energy) of $SrRhO_3$, $SrMoO_3$, $SrTiO_3$ and $SrZrO_3$ are 25 eV, 25.3 eV, 26 eV and 24.9 eV respectively since the highest peak at a certain frequency (light energy) related to the bulk plasma frequency.

The ratio between the energy of a wave reflected from a surface to the energy of the wave incident upon it is generally referred to as reflectivity. The reflectivity spectra of $SrTMO_3$ (TM = Rh, Ti, Mo, Zr) as a function of photon energy is shown in Fig. 8 (e) according to which the reflectivity is 0.30 – 0.34 in the infrared region for $SrRhO_3$, 0.18 – 0.19 for $SrMoO_3$, 0.19 – 0.21 for $SrTiO_3$ and 0.12 – 0.14 for $SrZrO_3$ respectively. These values drop rapidly in the high energy region with some peaks as a result of intraband transition [35].

The refractive indices (both the real and imaginary parts) of $SrTMO_3$ (TM = Rh, Ti, Mo, Zr) are illustrated in Fig. 8 (f). The imaginary part of refractive index indicates the amount of absorption loss of electromagnetic wave when propagates through the materials while the real part indicates the phase velocity of electromagnetic wave. We observed from the figure that the overall features of the refractive spectra for the all four oxide metals are same in the whole energy range without some variation in heights and positions of the peaks. The static values of the refractive indices are 3.1, 2.1, 2.5 and 2 for $SrRhO_3$, $SrMoO_3$, $SrTiO_3$ and $SrZrO_3$ respectively.

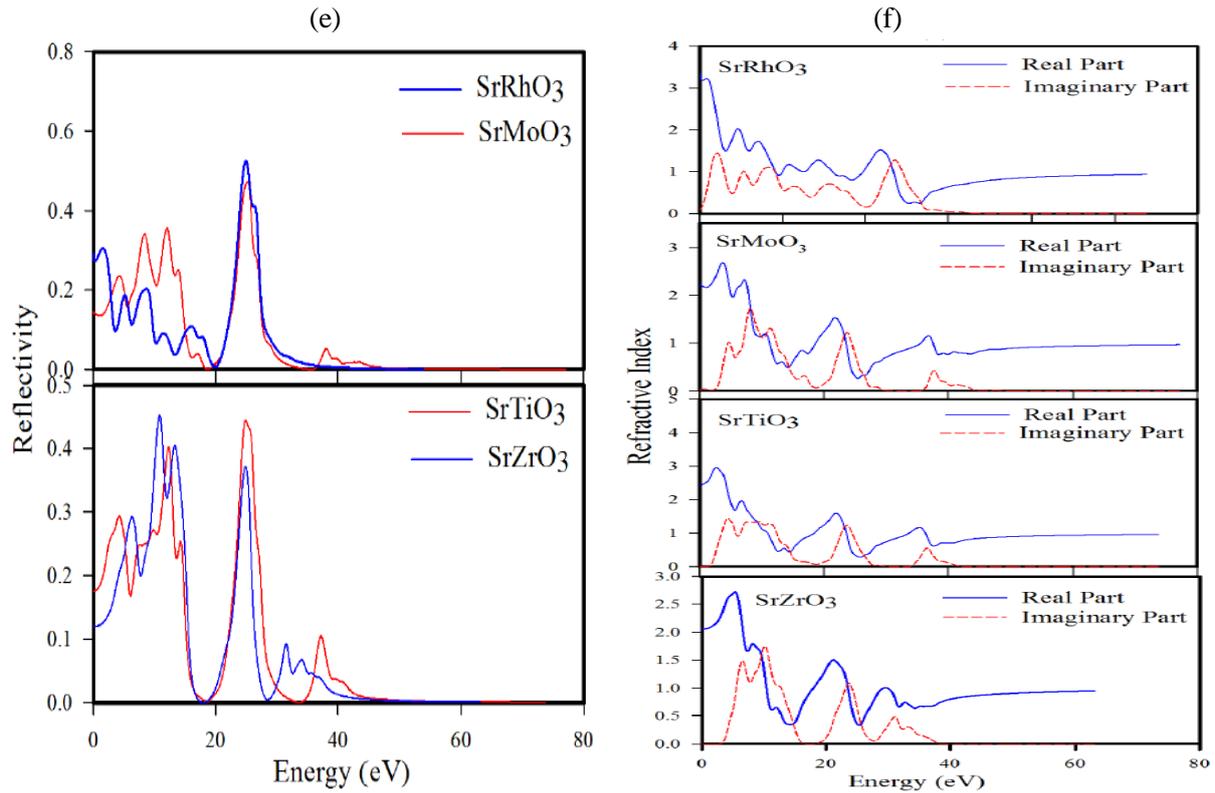

**Fig. 8.** The Reflectivity (e), and Refractive index (f) of SrTMO$_3$ (TM = Rh, Ti, Mo, Zr).

### 4. Conclusions

In summary, we have performed first principles calculations to investigate the structural, elastic and optical properties of SrTMO$_3$ within the framework of DFT. The calculated lattice parameters are in good consistent with the experimental and other theoretical results. We have studied the pressure dependence of elastic properties of SrTMO$_3$ compounds. Our calculated results show that the elastic constants $C_{11}$, $C_{12}$ and $C_{44}$ of SrTMO$_3$ almost increases with the increasing pressure in the range of 0 – 60 GPa with a step of 20 GPa. In addition, the bulk modulus, shear modulus, Young's modulus are also presented and discussed. By analyzing *B/G* and Poisson's ratios in various pressures, we find that SrTMO$_3$ compounds become more ductile with the increase in pressure. The pressure dependence of anisotropic factor are also calculated and analyzed. The investigation on the optical properties of SrTMO$_3$ (TM = Rh, Ti, Mo, Zr) metal oxides reveal that all of the four compounds under study possess good conductivity and absorption coefficient. According to the value of the dielectric constant SrRhO$_3$ can be used as a good dielectric material as the static dielectric constant of this compound is higher than the three others. In conclusion, it is expected that this theoretical investigation on the four metal oxide SrTMO$_3$ (TM = Rh, Ti, Mo, Zr) will help to use these materials for a wide range of practical application.

### References

1. J.B. Goodenough, Rep. Prog. Phys. 67 (2004) 1915.
2. K.Y. Hong, S.H. Kim, Y.J. Heo, Y.U.Kwon, Solid State Commun. 123 (2002) 305.
3. S. Takeno, T. Ohara, K. Sano, T. Kawakubo, Surf. Interface Anal. 35 (2003) 29.
4. H.H. Kung, Transition Metal Oxides: Surface Chemistry and Catalysis, Elsevier Science, NY, (1989).
5. I.R. Shen, V.L. Kozhevnikov and A.L. Ivanovskii, J. Solid State Science, Vol. 10, No. 2, (2008), p- 217-225.

6. A.J. Smith and A.J.E. Welch, J. Acta Crystallographica, Vol.13,(1960), p-653-656.
7. P. Hohenberg, W. Kohn, Phys. Rev. 136 (1964) B864–B871.
8. Materials Studio CASTEP manual_Accelrys, 2010. pp. 261–262. <http:// www.tcm.phy.cam.ac.uk/castep/documentation/WebHelp/CASTEP.html>.
9. J.P. Perdew, K. Burke, M. Ernzerhoff, Phys. Rev. Lett. 77 (1996) 3865.
10. S.B. Fagan, R. Mota, R.J. Baierle, G. Paiva, A.J.R. da Silva, A. Fazzio, J. Mol. Struct. 539 (2001) 101.
11. H. J. Monkhorst and J. D. Pack, Phys. Rev. B 13, 5188 (1976).
12. B. G. Pfrommer, M. Cote, S. G. Louie, and M. L. Cohen, J. Comput. Phys. 131, 233 (1997).
13. Fan CZ, Zeng SY, Li LX, Zhan ZJ, Liu RP, et al. Physical Review B 2006;74:125118-23.
14. E. Mete, R. Shaltaf, and S. Elliatioglu, Phys. Rev. Bm Vol. 63, No. 3, 2003, Article ID 0304703.
15. T. Mitsui and S. Nomura, Springer-Verlag, Berlin, 1982.
16. A. Daga, S. Sharma, K.S. Sharma, J. Mod. Phys. 2 (2011) 812-816.
17. F. Birch, Phys. Rev. 71 (1947) 809.
18. J. Y. Wang and Y. C. Zhou, Phys. Rev. B 69, 214111 (2004).
19. M.M. Wu, L. Wen, B. Y. Tang, L.M. Peng, and W. J. Ding, Journal of Alloys Compounds 506, 412 (2010).
20. Z. W. Huang, Y. H. Zhao, H. Hou, and P. D. Han, Physica B 407, 1075 (2012)
21. W. Voigt, Lehrbuch de Kristallphysik, Terubner, Leipzig (1928).
22. A. Reuss, Z. Angew. Math. Mech. 9, 49 (1929).
23. Y. Liu, W.-C. Hu, D.-J. Li, X.-Q. Zeng, C.-S. Xu, and X.-J. Yang, Intermetallics 31, 257 (2012).
24. R.D. King-Smith, D. Vanderbilt, Phys. Rev. B 49 (1994) 5828.
25. L. Bornstein, T. Mitsui, S. Nomura, Crystal and Solid State Physics, vol. 16, Springer-Verlag, Berlin (1982).
26. R. Terki, H. Feraoun, G. Bertrand, H. Aourag, Phys. Status Solidi B 242 (2005) 1054.
27. D. M. Teter, MRS Bull. 23, 22 (1998).
28. S.F. Pugh, Philos. Mag. 45 (1954) 823.
29. Y. Cao, J. C. Zhu, Y. Liu, Z. S. Nong, and Z. H. Lai, Comput. Mater. Sci. 69, 40 (2013).
30. I.N. Frantsevich, F.F. Voronov, S.A. Bokuta, in: I.N. Frantsevich (Ed.), Elastic Constants and Elastic Moduli of Metals and Insulators Handbook, 1990, Naukova Dumka, Kiev, 1982, pp. 60-180.
31. C. Zener, Elasticity and Anelasticity of Metals, University of Chicago Press, Chicago (1948).
32. P. Ravindran, L. Fast, P. A. Korzhavyi, and B. Johansson, J. Appl. Phys. 84, 4891 (1998).
33. J. Sun, X.F. Zhou, Y.X. Fan, J. Chen, H.T. Wang, Phys. Rev. B 73 (2006) 045108–045110.
34. Rahman, Atikur, Afjalur Rahman, and Zahidur Rahaman. "First-principles calculations of structural, electronic and optical properties of HfZn2." *Journal of Advanced Physics* 5.4 (2016): 354-358.
35. Rahman, Md, and Md Rahaman. "The structural, elastic, electronic and optical properties of MgCu under pressure: A first-principles study." *arXiv preprint arXiv:1510.02020* (2015).